# The surface charge of a cell lipid membrane

M. Pekker [1] and M. N. Shneider [2]

[1] MMSolution, 6808 Walker Str., Philadelphia, PA 19135, USA

[2] Department of Mechanical and Aerospace Engineering, Princeton University, Princeton, NJ 08544, USA

E-mail: pekkerm@gmail.com and m.n.shneider@gmail.com

**Abstract**

*In this paper the problem of surface charge of the lipid membrane immersed in the physiological solution is considered. It is shown that both side of the bilayer phospholipid membrane surface are negatively charged. A self-consistent model of the potential in solution is developed, and a stationary charge density on the membrane surface is found. It is shown that the ions of the surface charge are in a relatively deep (as compared to $k_BT$) potential wells, which are localized near the dipole heads of phospholipid membrane. It makes impossible for ions to slip along the membrane surface. Simple experiments for verifying the correctness of the considered model are proposed. A developed approach can be used for estimations of the surface charges on the outer and inner membrane of the cell.*

Introduction

The question about the distribution of electric charge near the cell membranes is the key for many problems associated with the interaction of cells with external electromagnetic fields [1-6]. For example, if ions are not electrically neutral, the areas are not strongly bound with the membrane and can move freely along it, the electromagnetic field has a weak influence on the membrane. If the ions are firmly bound to the membrane, the electrical component of electromagnetic fields could lead to the membrane deformation [7, 8].

The surface charge of the cell membrane has been studied in many experimental works. We will not consider all works and may refer only to [9,10], where the experimental data of the surface charge of cell membranes obtained by different methods are given. A range of results is broad enough: $\sigma_m = 0.3 - 0.002 \, C/m^2$.

The problem of spatial distribution of charge in the vicinity of the biological membranes surface has been considered in many теоретических papers, see eg [1-5]. In all these works, the near-surface potential of the membrane was considered under the Gouy-Chapman theory [11, 12] or its later modification by Stern [13], in which the charge on the membrane surfaces is considered to be given. In these theories, the membrane was considered as a continuous dielectric, without taking into account its fine structure, and a surface charge was determined on the basis of the electrochemical properties of the dielectric surface (see, eg [14, 15]).

In the classical Gouy-Chapman-Stern theory the surface charge ions can freely slide along the surface. Because, it is assumed in this theory, that the interaction of ions with the surface occurs due to electrostatic forces with the induced charges and absorption forces, which depends only on the distance of the ion to the surface. In the case of a phospholipid membrane, the electrolyte ions interact with phospholipid dipoles, which form a mosaic (lattice) structure (Fig.1a) [16]. As a result, the surface charge ions are in a periodic potential (Ris.1b), i.e. sliding them along the surface of the membrane is suppressed. This fact is crucial in the theory of interactions of the weak microwave field with the cell membranes, proposed in [17], where it was assumed that the longitudinal electric field of the microwave acting on the surface charge causes forced longitudinal vibrations of the membrane, because the surface charge ions are rigidly bonded to the surface and cannot slip.

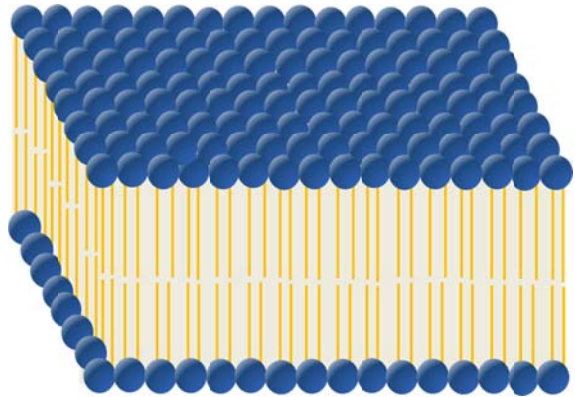

**Fig. 1a**. A simplified mosaic model of a phospholipid membrane [16]. Spheres represent the dipole heads on the inside and the outside of the membrane.

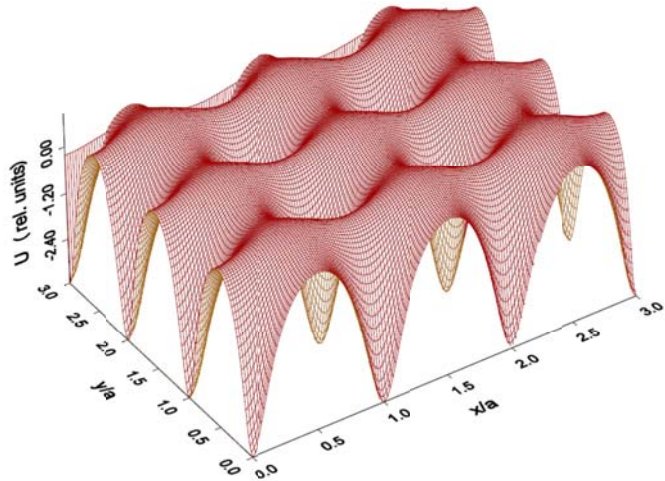

**Fig. 1b**. A fragment of a two-dimensional distribution pattern of potential along membrane surface, at a distance equal to 0.15 of the lattice period (a description of the computational model is given in Parts 1 and 2).

In our paper, we take into account the fine structure of the bilayer phospholipid membranes [16]. It is shown that the surface of the cell membrane is charged negatively with ions, trapped in the potential wells formed by the dipole heads of the membrane phospholipids. These ions are strongly bound to the membrane surface: the ion binding energy U with the membrane surface is much greater than the thermal energy $k_BT$. That fact allowed the construction of a self-consistent theory of Gouy-Chapman and Stern theories, and helped to determine the average charge density of sitting on the membrane.

The membrane surface charge theory presented in this paper supports the idea [17] of electromechanical impact of weak microwave electromagnetic fields on the surface density redistribution of transmembrane ion channels.

We emphasize that this article does not discuss the passage of ions through the transmembrane ion channels or pores. We restrict ourselves to considering only the surface processes in the lipid membrane placed in the salt-water solution.

The main ions in the physiological solution are: Na$^+$, K$^+$, Ca$^{2+}$, Mg$^{2+}$ and Cl$^-$. The total concentration of these: $n_i \approx 1.8 \cdot 10^{26}$ m$^{-3}$ (300mmL/Mole) [21], но для простоты мы рассмотрим только случай, когда the bilayer phospholipid membranes immersed in water solution of NaCl.

1. **A model of the phospholipid membranes.**

The dielectric permittivities of the membrane and water are correspondingly: $\varepsilon_m \approx 2$ and $\varepsilon_w \approx 80$ [18, 19]. It would seem that in order to evaluate the force acting on the ion of charge q near the membrane surface the formula [20] can be applied:

$$F_q = \frac{q^2}{4h^2} \frac{\varepsilon_w - \varepsilon_m}{\varepsilon_0(\varepsilon_w + \varepsilon_m)} \tag{1}$$

Here $\varepsilon_0$ is the permittivity of free space, and $h$ is the distance to the interface between dielectrics. However, the paradoxical result follows from the formula (1) that the ion cannot get from the dielectric with the higher dielectric permittivity into the dielectric with the lower dielectric permittivity, because the force acting on it increases inversely proportional to the square of the distance to the boundary between dielectrics. This statement contradicts the experimental facts and theoretical models of Hodgkin and Huxley's kind, based on the phospholipid membrane permeability to ions. Usually, in theories of the interaction of ions with the surface of dielectrics, the minimum value of $h$ is selected as the size of the ion, or the Debye radius, if the insulator's surface is saturated with ions [15]. However, this does not resolve the problem of phospholipid membranes "impenetrability". In fact, the problem lies in that the formula (1) is derived in the approximation of continuous medium with the dipoles of the infinitely small size, without taking into account the real structure of the surface layer of the membrane.

It is known that the phospholipid molecules of cell membrane are forming a mosaic (matrix) structure in which dipole heads are directed towards the liquid (positively charged head faces outward membrane) [16]. The average surface area per molecule of the lipid is $\approx 0.5$ nm$^2$, the length of the polar head is $\sim 0.5 - 1$ nm, the radius of the head is $\sim 0.2 - 0.3$ nm, and the distance between the hydrophilic heads of the membrane is in the range of 5-7 nm [21, 22]. The dipole moment of the phospholipid head is 18.5-25 D [23] (1 D $= 3.34 \cdot 10^{-30}$ C·m), i.e. more than 10 times greater than the dipole moment of water molecules. On the basis of geometrical dimensions of the cell membrane and the size of water molecules, it can be concluded that the free space between the head does not exceed the size of a water molecule (~ 0.2nm). That is, the membrane, interacting with the ions of surrounding liquid, cannot be considered as a dielectric medium with an infinitesimal dipoles size.

These facts allow the consideration of the following simplified model of the ion interaction with the membrane:
1. The membrane represents a matrix (Fig. 2) with a mesh size $a \times a$. In the nodes of cells the dipoles are located; the dipole charge is q; the distance between the charges (the dipole length), d; the distance between the dipoles along the axis z, l.
2. An ion is a classical particle and cannot approach a dipole at a distance less than the sum of the radii of the head and the size of the ion. It is important that the ion can approach the membrane dipole heads close enough that would be "captured" by the potential well. This is a standard assumption in the theory of the interaction of ions with the surface of the dielectric [15]. Since the dipole moment of water is 10 times less than the dipole moment of the phospholipid head molecule and near the head can not be more than one-two water molecules, the interaction between the ions located near the surface of the membrane and the water molecules can be neglected, as compared with the interaction of the ions with the dipoles of the membrane.
3. We want to emphasize one more time that ionic permeability membrane is a separate problem and not consider in this work, for us it is important to know how strong ions attached to surface of membrane.

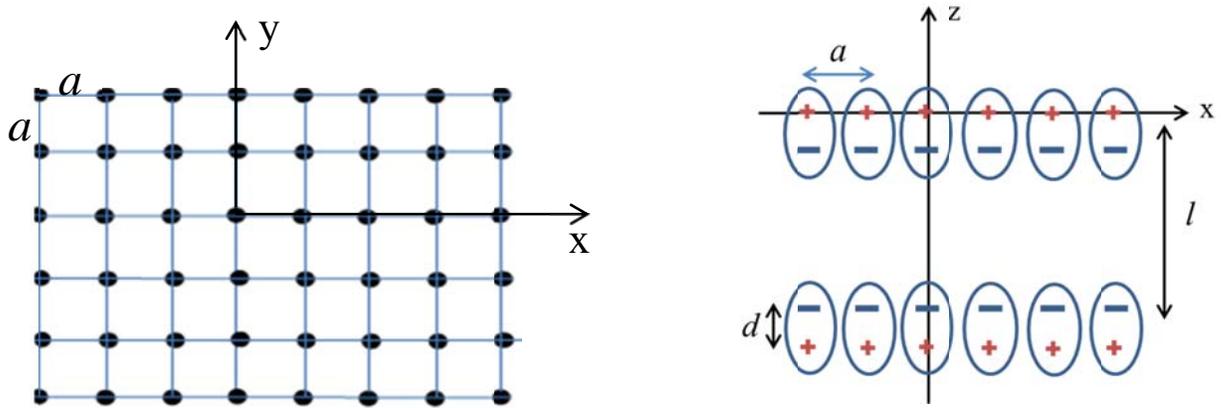

**Fig. 2**. A simplified mosaic model of a membrane: the membrane is a matrix with dipoles in the nodes. $a$ is the mesh size of the matrix, $d$ is the distance between the charges of the dipole head, $l$ is the distance between the dipoles along the z axis.

The main ions in the physiological solution are: $Na^+$, $K^+$, $Ca^{2+}$, $Mg^{2+}$ and $Cl^-$. The total concentration of these: $n_i \approx 1.8 \cdot 10^{26}$ m$^{-3}$ (300mmL/Mole) [21], but for simplicity we consider only the case when the bilayer phospholipid membranes are immersed into the water solution of NaCl.

## 2. The potential near the surface of the membrane

The expression for the membrane potential in a cylindrical coordinate system $(r, \theta, z)$ at a point above the surface of the membrane (Fig. 3):

$$U(r,\theta,z) = \sum_{i=-n}^{n} \sum_{j=-n}^{n} U_{i,j}(r,\theta,z) =$$

$$\sum_{i=-n}^{n} \sum_{j=-n}^{n} \frac{q}{4\pi\varepsilon_0} \left( \frac{1}{\sqrt{r^2 + r_{i,j}^2 - 2r \cdot r_{i,j} \cos(\theta - \theta_{i,j}) + z^2}} \right.$$

$$- \frac{1}{\sqrt{r^2 + r_{i,j}^2 - 2r \cdot r_{i,j} \cos(\theta - \theta_{i,j}) + (z+d)^2}} \quad (2)$$

$$- \frac{1}{\sqrt{r^2 + r_{i,j}^2 - 2r \cdot r_{i,j} \cos(\theta - \theta_{i,j}) + (z+l)^2}}$$

$$\left. + \frac{1}{\sqrt{r^2 + r_{i,j}^2 - 2r \cdot r_{i,j} \cos(\theta - \theta_{i,j}) + (z+l+d)^2}} \right)$$

Here, the radial and angular coordinates $r_{i,j}, \theta_{i,j}$ correspond to the position of the dipole in the node with the numbers ($i, j$), $q$ is the dipole charge, and $\varepsilon_0$ is permittivity of free space. The value of n in (2) is chosen such that the potential near the membrane does not depend on the size of the matrix n. The count goes from the node (dipole head) (Fig. 3): $r = (x^2 + y^2)^{1/2}$, $\theta = \arctan(y/x)$.

We expand the potential $U$ in a Fourier series by $\theta$. Taking into account the rotational symmetry on $\pi/2$ rotations for the accepted square matrix of the dipole heads, we obtain:

$$U(r,\theta,z) = U_0(r,z) + \sum_{j=1}^{J} U_j(r,z)\cos(4j\theta) + \sum_{j=1}^{J} V_p(r,z)\sin(4j\theta)$$

$$U_0(r,z) = \frac{2}{\pi} \int_0^{\pi/2} U(r,\theta,z)d\theta \qquad V_0(r,z) = 0 \qquad (3)$$

$$U_{j\neq 0}(r,z) = \frac{4}{\pi} \int_0^{\pi/2} U(r,\theta,z)\cos(4j\theta)d\theta \qquad V_{j\neq 0}(r,z) = \frac{4}{\pi} \int_0^{\pi/2} U(r,\theta,z)\sin(4j\theta)d\theta$$

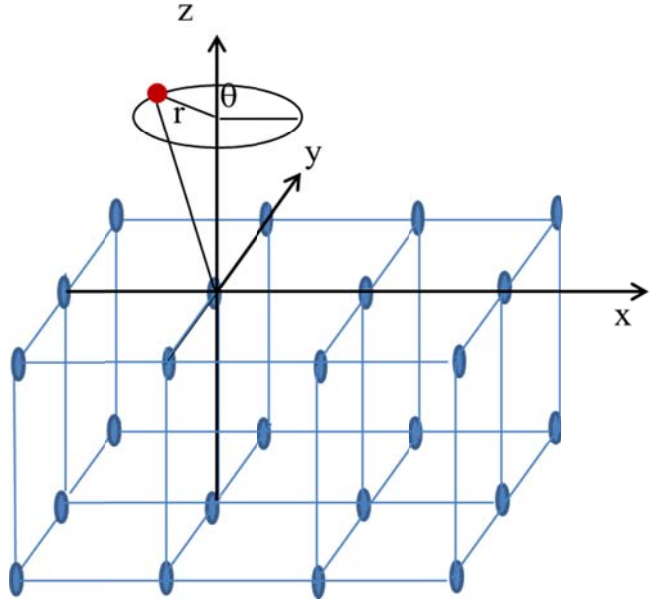

**Fig.3.** Spatial lattice of dipoles in bilayer lipid membrane. Potential is calculated at the point P. Ellipsoids at lattice sites correspond to the dipole heads of phospholipid molecules.

In the framework of the model presented above, estimates show that the influence of the dipoles located on the lower side of membrane (Fig. 2) on the potential distribution $U$ at $z > 0$ can be neglected. The same is true to the effect of dipoles in the upper side of the membrane on the potential at $z < -(l+d)$. As far as the dependence of $U$ on $\theta$, in the region $r \leq a/2 = 0.35$ nm the contribution of above the zero harmonics does not exceed 7%, so we will henceforth neglect them.

For example, let consider the case, when bilayer phospholipid membrane is immersed in the NaCl solution, at the following set of parameters: $a = 0.7$ nm, $d = 0.5$ nm, $l = 8$ nm, and $q = 1.6 \cdot 10^{-19}$ C [16, 21].

Fig. 4. shows contour plot of reduced $U/k_B T$ for a singly charged ion at $T = 300$K, Fig. 5 the dependence $U/k_B T$ on the axis $z$.

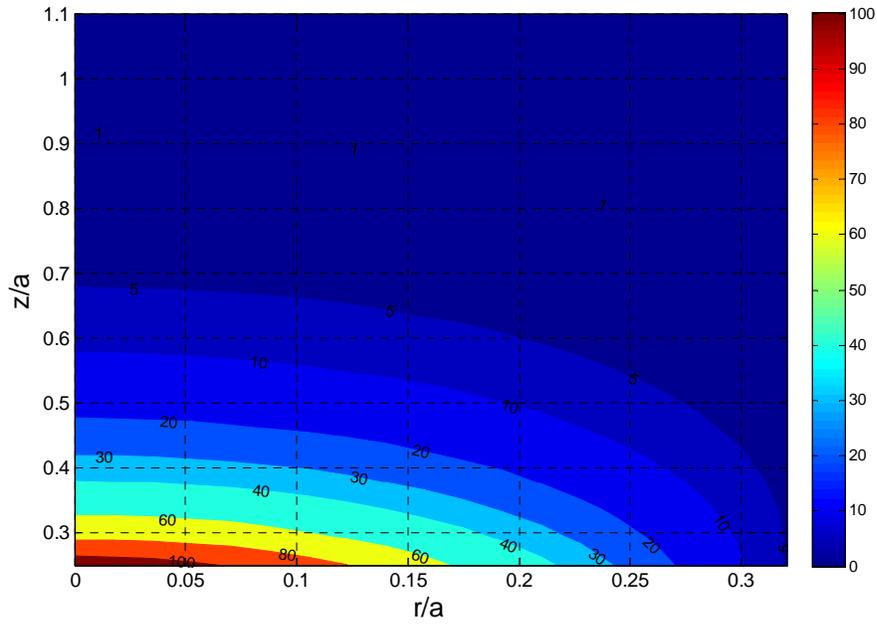

**Fig. 4**. Contour plot $U/k_B T$, corresponding to a half of the cell shown in Fig. 1b.

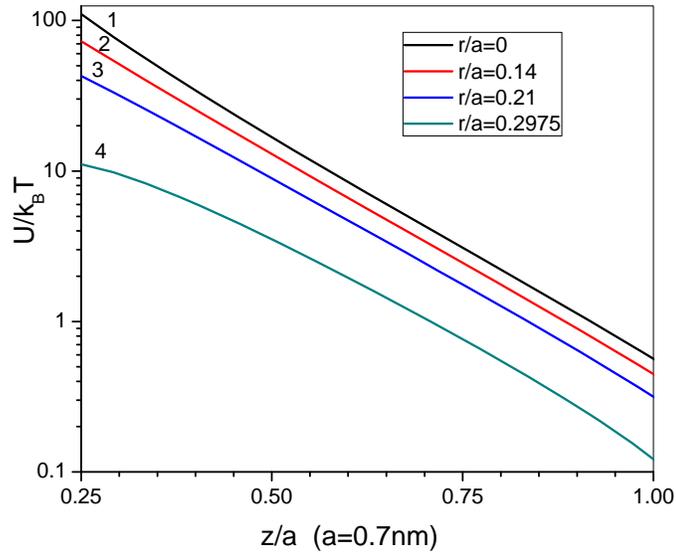

**Fig. 5.** The parameter $U/k_B T$ dependence on z. Curve 1 corresponds to $r/a = 0$, 2 – $r/a = 0.14$, 3 - $r/a = 0.21$, and 4 – $r/a = 0.30$.

It is seen from Fig, 5 that the value $U/k_B T$ decreases exponentially with the distance from the membrane, wherein a distance $z/a = 0.5$, the ratio $U/k_B T$ is about 10. That is still sufficiently deep potential well, but already at $z/a = 1$, the influence of the membrane on the ions can be neglected.

An effective potential of the ion across the membrane can be determined. Let the ion be a sphere of radius $r_i$, and $r_h$ is the distance from the positive charge of the dipole to the head of the ion (Fig. 6).

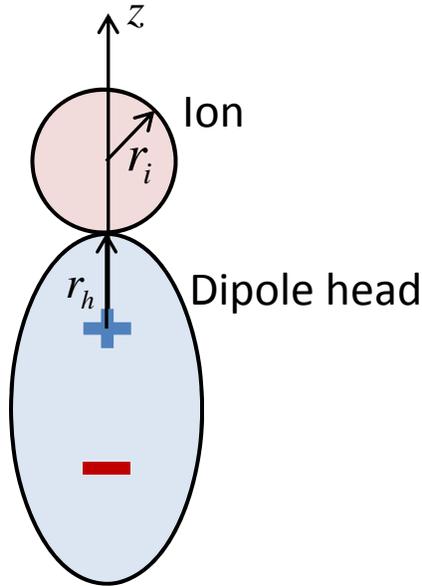

**Fig. 6** The negative ion located near the head of the phospholipid molecule dipole. Ion is a charged sphere of radius $r_i$, and $r_h$ is the distance from the positive charge of the dipole to the head of the ion.

In this case, the effective reduced ion potential,

$$\frac{U_{ef}}{k_B T} = \frac{2\pi}{k_B T} \frac{3}{4\pi r_i^3} \int_{r_h}^{r_h+2r_i} dz \int_0^{\sqrt{r_i^2-(r_h+r_i-z)^2}} U(r,z) r\, dr\, dz \qquad (4)$$

The dependence of the effective potential of the ion $Cl^-$ in the potential well on the size of the head $r_h$ is shown in Figure 6. The chlorine ion radius $r_i = 0.18$ nm, and the atomic weight is $M_i = 35.4$.

It is seen that for the typical size of the dipole head $r_h = 0.2$ nm ($r_h/a = 0.29$) and the ion radius $r_i = 0.18$ nm, the reduced potential depth is $U_{ef}/k_B T \approx 10$. Thus, the trapped ion is quite firmly bound to the membrane. It should be noted that for electrolytes of a different composition of negative ions, for example in the axon, where the major negative ions are anion groups of macromolecules and phosphates, estimations of the potential will differ from that shown in Fig.7.

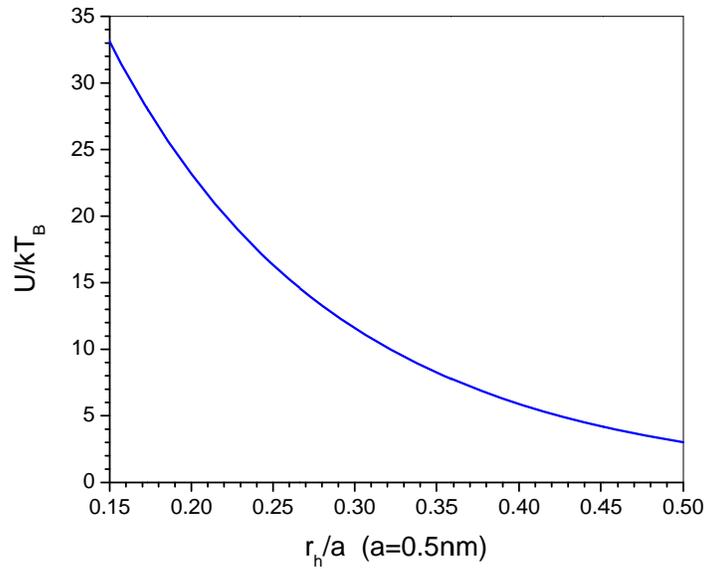

**Fig. 7.** The dependence of the effective potential of the ion on the size of the head at the assumed radius of the ion $r_i = 0.18$ nm (Cl⁻).

### 3. Evaluation of the membrane surface charge.

For simplicity, we assume the ions as a point charges, but will take into account the finite size of the ion as the minimal distance that the ion can approach to the membrane. As indicated above, this is a standard approach in models of the ions interaction with the dielectric surfaces.

We assume that the space charge on the membrane does not affect its structure, i.e. the dipole heads. Since the field of the membrane (without taking into account the ionic charge on it) at a distance of the order of mesh size is negligibly small, and the distance between the free ions in the solution is larger than the mesh size, we can relate the charge on the membrane surface to the charges in solution by equating the flows at the membrane/liquid boundary (Fig. 8).

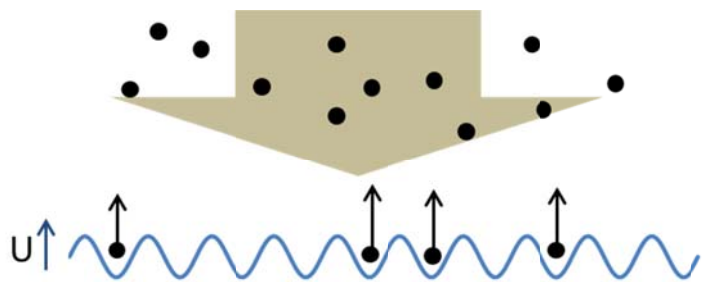

**Fig. 8.** A scheme of the ions flows near the membrane. The ions flux to the membrane and "evaporation" of the ions from the potential wells of the depth U.

Let us estimate the flux of ions falling on the membrane. Assuming the Boltzmann distribution of ions outside the membrane:

$$n = n_\infty \exp\left(-\frac{q\varphi}{k_B T}\right), \qquad (5)$$

where $n_\infty$ - is ion density at infinity. Assuming that the ions are in thermal equilibrium with the water molecules, we obtain the following estimate for the ion flux on the surface of the membrane:

$$P_{in} = \frac{1}{2\sqrt{\pi}} n_\infty v_{H_2O} \left(\frac{M_{H_2O}}{M_i}\right)^{1/2} \exp\left(-\frac{q\varphi}{k_B T}\right). \qquad (6)$$

Here $v_{H_2O}\left(\frac{M_{H_2O}}{M_i}\right)^{1/2}$ is the averaged ion velocities in solution, and $M_{H_2O}, M_i$ are the masses of the water molecule and the ion, correspondingly.

On the other hand, an estimate of the ion flux "evaporating" from the surface of the membrane is as follows. The flow of ions from the surface of the membrane can be considered as a process of evaporation of the liquid with a work function equal to the potential well $U$. The ion in the well is exposed to random impacts by water molecules, while the ion's presence time in the potential well $\tau_i$ can be estimated as:

$$\tau_i \approx \tau_{i,H_2O}\left(\frac{M_i}{M_{H_2O}}\right)(\exp(U/kT)-1). \qquad (7)$$

Here $\tau_{i,H_2O}(M_i/M_{H_2O})$ is the characteristic time of energy exchange between the ion and the water molecules. In (7) we have taken into account that at $U/kT \ll 1$ the ions are reflected from the membrane, i.e., the time delay is zero. For estimates $\tau_{i,H_2O}$ can be assumed equal to the collision time of water molecules in a solution $\tau_{i,H_2O} = \lambda_{H_2O}/v_T$, $\lambda_{H_2O} = 0.2 \cdot 10^{-9}$ nm is the mean free path of a molecule of water, and $v_T$ is the thermal velocity of water molecules. Accordingly, the number of the ions leaving the unit surface of the membrane per unit time is equal to:

$$P_{out} = \frac{1}{a^2}\frac{N_i}{\tau_i} = \frac{N_i v_{H_2O}}{a^2 \cdot \lambda_{H_2O}}\left(\frac{M_{H_2O}}{M_i}\right)\frac{1}{\exp(U/k_B T)-1}. \qquad (8)$$

Here $1/a^2$ is the number of potential wells per unit membrane surface; $N_i$ is the relative population of the potential wells with ions. Equating the incident flux of the ions on the membrane to the "evaporating" flow from it, we obtain the relative population of the potential wells:

$$\frac{N_i}{1-N_i} = \frac{1}{2\sqrt{\pi}} n_\infty a^2 \cdot \lambda_{H_2O} \left(\frac{M_i}{M_{H_2O}}\right)^{1/2} \exp\left(-\frac{e\varphi_0}{k_B T}\right)\left(\exp\left(\frac{U}{k_B T}\right)-1\right). \qquad (9)$$

In (8) we took into account that the probability of the incident ion being captured for in a potential well is proportional to $1-N_i$.

Now we will find the distribution of electric potential in the liquid. Following the Debye approximation for the liquid electrolyte [11]:

$$\varphi = \varphi_0 \exp(-x/\lambda_D) \qquad (10)$$

Where $\lambda_D = \sqrt{\dfrac{\varepsilon \varepsilon_0 k_B T}{2 n_\infty q^2}}$ is the Debye length.

The field on the membrane surface:

$$\left.\dfrac{\partial \varphi}{\partial x}\right|_{x=0} = \dfrac{\sigma}{2\varepsilon \varepsilon_0} \qquad (11)$$

Where $\sigma$ is the surface charge density. For the case when the upper and lower surfaces of the membrane are equally charged $\sigma = 2 N_i q / a^2$, we obtain the potential at the membrane surface:

$$\varphi_0 = \dfrac{\lambda_D}{\varepsilon \varepsilon_0} \dfrac{q \cdot N_i}{a^2} = \sqrt{\dfrac{k_B T}{2 \varepsilon \varepsilon_0 n_\infty}} \cdot \dfrac{N_i}{a^2} \qquad (12)$$

Substituting (12) into (9), we obtain the equation for the value of $N_i$:

$$\dfrac{N_i}{1-N_i} = \dfrac{1}{2\sqrt{\pi}} n_\infty a^2 \cdot \lambda_{H_2O} \left(\dfrac{M_i}{M_{H_2O}}\right)^{1/2} \exp(-\eta N_i) \cdot \left(\exp\left(\dfrac{U}{k_B T}\right) - 1\right)$$

$$\eta = \dfrac{q}{a^2}\sqrt{\dfrac{1}{2\varepsilon\varepsilon_0 n_\infty k_B T}} \qquad (13)$$

For parameter values that are typical for a physiological solution: $n_\infty = 0.9 \cdot 10^{26}\,\text{m}^{-3}$, $a^2 = 0.5 \cdot 10^{-18}\,\text{m}^2$, $T = 300K$, $\lambda_{H_2O} = 0.2$ nm, for the chloride ion $(M_i / M_{H_2O})^{1/2} \approx 1.41$ we obtain from (13) $\eta \approx 13.9$, and, therefore,

$$\dfrac{N_i}{1-N_i} = 0.025 \cdot (\exp(U/k_B T) - 1) \cdot \exp(-13.9 \cdot N_i) \qquad (14)$$

The calculated dependence of $N_i$ on the reduced potential $\dfrac{U}{k_B T}$ is shown in Figure 9.

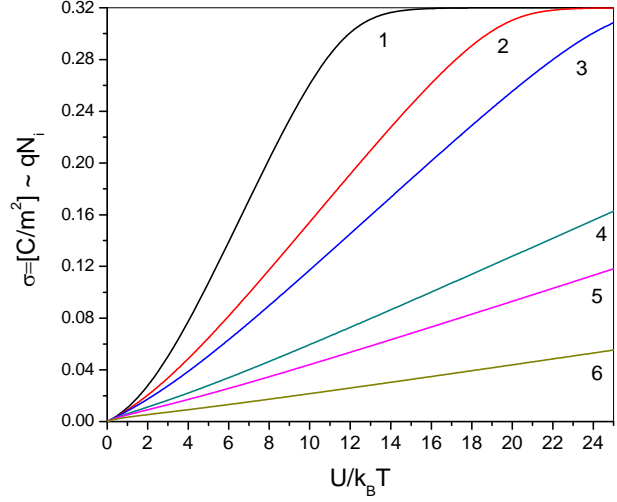

**Fig. 9.** Dependencies of the relative occupancy of the membrane $N_i$ and the surface charge density $\sigma$ on the reduced potential $U/k_BT$. Curve 1 corresponds to $n_\infty = 5\cdot 10^{26}$, 2 – $n_\infty = 10^{26}$, 3 – $n_\infty = 5\cdot 10^{25}$, 4 – $n_\infty = 10^{25}$, 5 – $n_\infty = 5\cdot 10^{24}$, 6 – $n_\infty = 10^{24}$.

The maximum surface charge density in Fig. 9 corresponds to the case where all of the cells (Fig. 1b) are full, i.e. one negative ion is in every cell. It is seen that, even for small values $U/k_BT$, the relative occupancy of the potential wells with charges is large enough and the charge density of the charges, which are tightly bounded to the membrane, can reach hundredths of coulomb per square meter. It should be noted that in this estimation we did not consider the interaction between the ions bonded to the membrane, which can certainly play an important role for a more accurate calculation of the occupancy of the potential wells.

Let us estimate the charge of the free ions of one sign in the non-quasineutral region in the vicinity of the membrane surface.

$$q_{free} \approx q \cdot n_\infty \cdot r_D \approx 0.013 C/m^2 \qquad (15)$$

For typical values of the potential $U/k_BT \approx 5-15$, which correspond to the size of the head $r_h \sim 0.2-0.3$ nm (Fig.9), a bounded charge sitting on the membrane surface is much larger than the charge associated with unbounded ions.

## 4. Discussion

In this article, we examined the behavior of the membrane surface with the dipole heads placed in an electrolyte solution. However, despite of this very simple approach, the results are directly related to the processes in biological lipid membranes, such as the membranes of the axon. Because the negative ions inside of the axon differ from those located on the outside (inside the main negative ions are anion groups of macromolecules and phosphates, but chloride ions outside), it is expected that the surface charge density of ions sitting on the membrane is different. Figure 10 shows the potential distribution inside and outside of the membrane where the inner and outer sides of the membrane have the same (A) and different (B) surface charges.

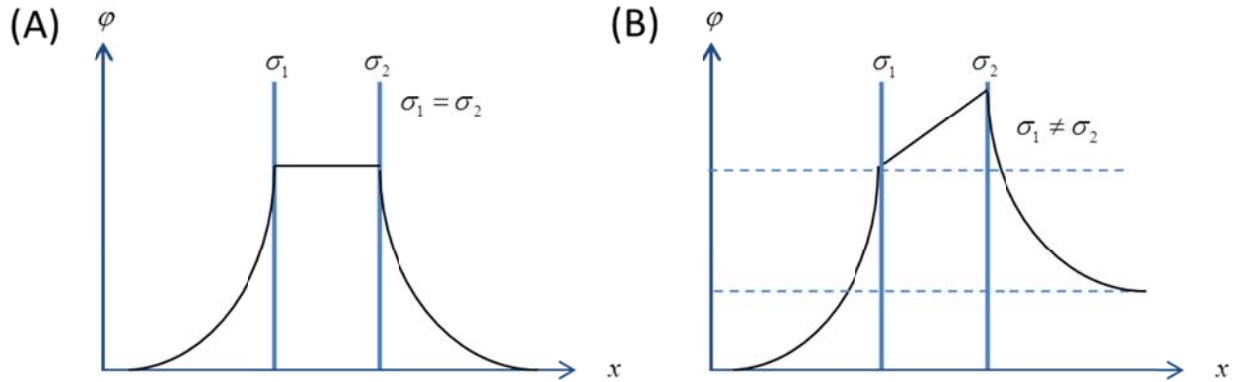

**Fig. 10.** Potential distribution inside and outside the membrane when the inner and outer sides of the membrane have the same (A) and different (B) surface charges.

A few words about the experimental verification of the proposed model. As follows from (13), the main parameters that determine the relative occupancy of the membrane with negative ions $N_i$, are $\eta \propto 1/\sqrt{N_\infty}$ and $U/k_B T$. Accordingly, by changing the density of NaCl in the solution and measuring the surface charge of the membrane, the curves, similar to those shown in Fig.8, can be reproduced. It is shown in [17] that the longitudinal displacements of the membrane in the microwave field is resonant and is linearly dependent on the surface charge density (Fig. 10). Accordingly, by changing the salinity of the water, it is possible to determine the surface charge of the membrane and to compare with estimates obtained above.

In [24,25], the distribution of ions near the surface of a phospholipid membrane was calculated using molecular dynamics, however, in those articles the binding energy of the ions with phospholipid heads was not illustrated. Therefore, it would be interesting to compare the corresponding results of our model and calculations performed by using the molecular dynamics.

**Conclusions**

It is shown in this paper that:

1. The effective binding energy of the ion with the membrane is of order of several $k_B T$. i.e., the ions are firmly bounded to the surface of the membrane.
2. The value of the potential decays exponentially with distance from the membrane, so that at a distance greater than 0.7 nm the influence on the ions of the electrolyte by the membrane can be neglected.
3. The surface charge ions are in relatively deep potential wells, localized near the dipole heads of the phospholipid membrane, which prevents them from sliding along the surface of the membrane.
4. A qualitative self-consistent theory of the potential distribution near the membrane is considered. It is shown that the density of the bound charges on the membrane can reach hundredths of C/m².
5. This work supports the assumption that the ions located on the excitable axon membrane are tightly bound with it, which underlies the work [17]. Therefore, the electric component of the microwave field, interacting with the ions, transfers energy and momentum directly to the membrane. This interaction leads to forced mechanical vibration of the membrane and, as a result, to a redistribution of transmembrane protein ionic channels.

It should be noted that the proposed model of the potential distribution near the membrane can be extended to for different types of biological membranes, taking into account their characteristics.

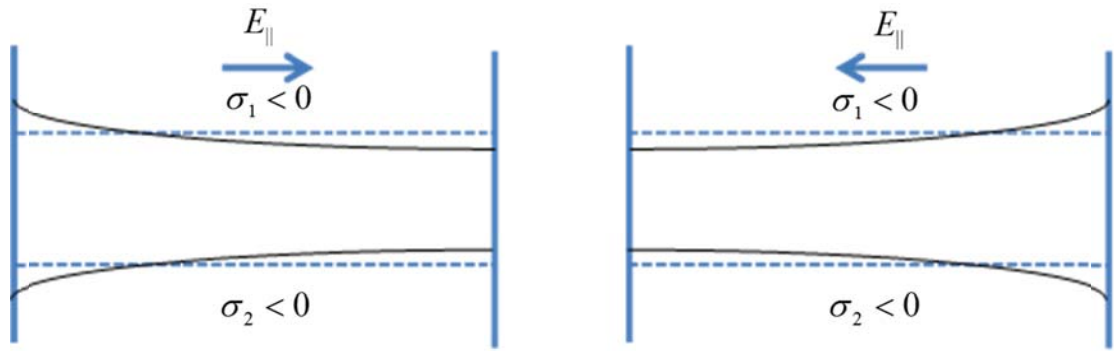

**Fig. 11.** Membrane deformations caused by the microwave electric field parallel to the surface of the membrane; $\sigma_1$, $\sigma_2$ are the inner and outer surface charge densities.